# Error estimation in the method of quasi-optimal weights

A.D. Morozov[1], A.V. Lokhov[2,3], F.V. Tkachov[3]

[1] Physics Department, Lomonosov Moscow State University, Moscow 119991, Russia
[2] University of Münster, Muenster 48149, Germany
[3] Institute for Nuclear Research of Russian Academy of Sciences, Moscow 117312, Russia

**Abstract**

We examine the problem of construction of confidence intervals within the basic single-parameter, single-iteration variation of the method of quasi-optimal weights. Two kinds of distortions of such intervals due to insufficiently large samples are examined, both allowing an analytical investigation. First, a criterion is developed for validity of the assumption of asymptotic normality together with a recipe for the corresponding corrections. Second, a method is derived to take into account the systematic shift of the confidence interval due to the non-linearity of the theoretical mean of the weight as a function of the parameter to be estimated. A numerical example illustrates the two corrections.

## 1. Introduction

Error estimation is an important and often difficult part of any physical experiment [1]. A key role is played by the concept of confidence interval: such interval must cover the unknown value being measured with a given probability called the confidence level.

In this work we study the problem of construction of confidence intervals for the method of quasi-optimal weights (QOW) [2], [3]. QOW is a rather general and versatile method of parameter estimation. It was successfully used to complete the data processing of the Troitsk-ν-mass experiment [4] after several years of difficulties that became known as the Troitsk anomaly [5]. The QOW method starts with the classical method of moments and inherits its analytical transparency (see textbooks [1], [6], [7], [8]). This allows one to approach the construction of confidence intervals in a systematic fashion, relying on analytical methods, which (together with its generality) distinguishes the QOW method from other approaches to parameter estimation including the maximum likelihood method. This is the more attractive that the QOW method is versatile and flexible (see examples of unconventional applications in [9]).

The present work is focused on the single parameter case as a starting point for any multi-parameter extensions.

The plan of our work is as follows.
In sec. 2 we review the method of quasi-optimal weights together with the basic large-$N$ formula for the confidence interval.
In sec. 3 we obtain a criterion for taking into account corrections that modify the confidence interval to account for insufficiently large sample size $N$.
Sec. 4 presents a method to correctly take into account the non-linearity of the function $h(\theta)$ (see the main equation of the QOW method, eq. (4)) which represents the theoretical mean of the quasi-optimal weight as a function of the parameter $\theta$ being estimated. It is shown that the ignoring of non-linearity can lead to a systematic general shift of the boundaries of the confidence interval, and also to an effective decrease of the confidence level.
In sec. 5 we discuss two basic examples: the Poisson case (sec. 5a) and the case of the normally distributed measurements (sec. 5b).
We always consider the case of a heterogeneous sample unless stated otherwise.

## 2. A review of the QOW method

The QOW method is a refinement of the classical method of moments in regard of the choice of the weight functions ("generalized moments"). Therefore, it is convenient to begin with a review of the classical method in order to fix notations in such way as to automatically account for the case of heterogeneous sample. A special attention is given to error estimation because this point is not treated in textbooks adequately.

The following notations will be used, following ref. [3]:

$\theta$ is a generic notation for the parameter being estimated, its various specific values are marked by indices: $\theta_0$, $\theta_i$ etc.;

$\theta^*$ is the unknown true value of the parameter which is to be estimated;

$x_i$ is the $i$-th random value representing one measurement;

$\{x_i\}$ is the sample of $N$ measurements;

$\pi_i(x;\theta)$ is the probability density function of $x_i$, it depends on $\theta$; this density is assumed to be known (analytically or as a Monte Carlo event generator; in the latter case it is sufficient to construct the main functions of the method $h(\theta)$ etc. in the form of numerical interpolations).

The method of moments assumes that one can compute theoretical means using the density distributions $\pi_i(x;\theta)$ for functions $f(x)$ of the random variable $x$ (the functions $f$ must satisfy the usual restrictions so that the emerging integrals make sense):

$$h_i(\theta) \triangleq \langle f_i \rangle_\theta \equiv \int dx\, \pi_i(x;\theta) f(x). \tag{1}$$

We will be calling $f$ weights rather than moments, retaining the latter word only in the name of the method, because we will not limit ourselves with the classical moments $f(x) = x^n$.

Besides the obvious requirement that the integral (1) exists, one may require continuity for each of the functions $f_i(x)$ [6]. For a convenient estimate of the dispersion to be possible, one should also require quadratic integrability of each $f_i$. In practice the weights in the QOW method satisfy, as a rule, both these requirements (even in the case of estimating the median of the Cauchy distribution that does not even have a mean).

It is convenient to introduce the following function (in all such expressions summations run over $i$ from 1 to $N$):

$$h(\theta) \equiv \langle f \rangle_\theta \triangleq \frac{1}{N} \sum_i \langle f_i \rangle_\theta \tag{2}$$

The corresponding empirical analogue (the empirical mean):

$$\langle f\rangle_{\exp} = \frac{1}{N}\sum_i f_i(x_i) \qquad (3)$$

The idea of the method of moments (inherited by the QOW method) is to take the solution with respect to $\theta$ of the following equation as an estimator for $\theta^*$ (see fig. 1):

$$\langle f\rangle_{\exp} = \langle f\rangle_\theta \equiv h(\theta) \qquad (4)$$

Within the adopted assumptions the empirical mean $\langle f\rangle_{\exp}$ tends to $h(\theta^*)$ as $N\to\infty$, and one expects that the solution of the equation with respect to $\theta$ would approach $\theta^*$. The corresponding (point) estimate is as follows (see fig. 1):

$$\theta_{\exp} = h^{-1}\left(\langle f\rangle_{\exp}\right). \qquad (5)$$

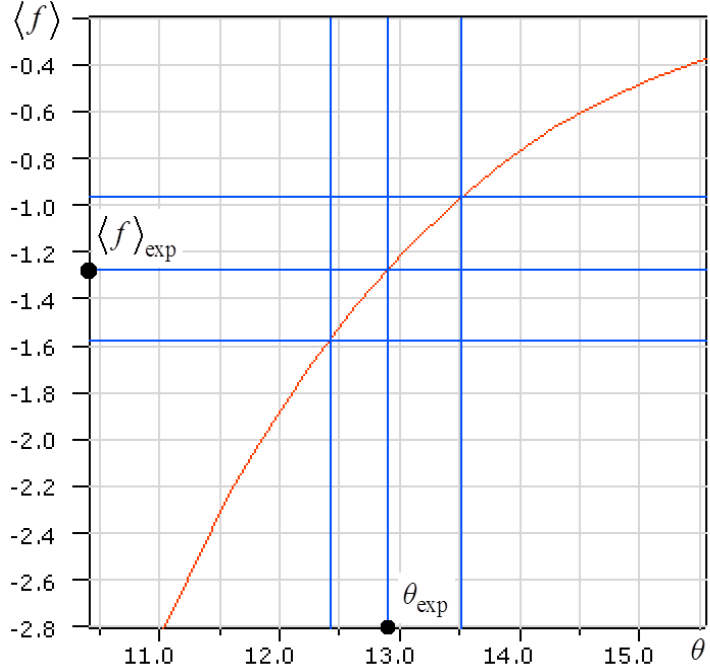

Fig. 1. The function $h(\theta)$ (the theoretical weight mean, eq. (2)) together with the point estimates and the CL= 90% intervals for the numerical example of sec. 5c.

It is clear from fig. 1, that the confidence interval for the estimate of $\theta$ in this method is obtained by mapping the confidence interval for $\langle f\rangle_{\rm th}$ using the inverse of $h$.

The estimation of $\langle f\rangle_{\rm th}$ is a special case of the so-called direct measurements [11]. Following the well-known recipe, take the unbiased sample variance

$$\mathrm{Var}_{\exp} f = \frac{N}{N-1}\left(\langle f^2\rangle_{\exp} - \langle f\rangle_{\exp}^2\right) \equiv \sigma_{\exp}^2. \qquad (6)$$

Then the confidence interval is written in a standard form:

$$f_\alpha^- \leq \langle f\rangle_{\rm th} \leq f_\alpha^+ \text{, где } f_\alpha^\pm = \langle f\rangle_{\exp} \pm \Delta f_\alpha,\ \Delta f_\alpha = y_\alpha N^{-1/2}\sigma_{\exp} \qquad (7)$$

where $y_\alpha$ is the $\alpha$-quantile of the standard normal distribution $N(0,1)$. One sees from fig. 1 that the interval (7) maps to

$$h^{-1}\left(f_\alpha^-\right) \leq \theta^* \leq h^{-1}\left(f_\alpha^+\right) \tag{8}$$

which is the required confidence interval for the confidence level $\alpha$ for the parameter being estimated. This formula is the starting point for the construction of confidence intervals in the QOW method.

Since the scatter of $\langle f \rangle_{\text{exp}}$ decreases for larger $N$, in practice (especially in the multidimensional case) it is convenient to have in view a linear approximation for this formula, perhaps together with quadratic corrections (see below).

The QOW method refines the basic method of moments by a special choice of the weights $f$. Within some natural assumptions on $\pi(x;\theta)$ there exist weights that generate estimates with the minimal possible variance (the so-called Frechet-Rao-Cramer limit [6]). The simplest one is given by the following formula

$$\varphi_{\text{opt}}(x;\theta) \triangleq \frac{\partial}{\partial \theta} \ln\left(\pi(x;\theta)\right) \tag{9}$$

and is called optimal weight (optimal, because the estimate obtained with this weight has the minimal variance). The optimal weight is defined up to an additive constant and a factor, both independent of the sample $\{x_i\}$.

For a heterogeneous sample, it is sufficient to consider the entire sample as a random variable whose probability distribution is a product of distributions for its components (due to assumed independence of different $x_i$), and then obtain the optimal weight according to the general prescription (9):

$$\begin{aligned}\varphi_{\text{opt}}(x) &\equiv \varphi_{\text{opt}}(x_1, x_2, \ldots, x_N) = \frac{\partial}{\partial \theta} \ln \prod_i \pi_i(x_i;\theta) = \sum_i \frac{\partial \ln \pi_i(x_i;\theta)}{\partial \theta} \\ &= \sum_i \varphi_{\text{opt},i}(x_i)\end{aligned} \tag{10}$$

In other words, the optimal weight for the entire sample is equal to the sum of optimal weights for individual measurements.

An important property of the optimal weight is the equality to zero of its theoretical mean when calculated over the distribution at the same value of the parameter, for which the weight is optimal:

$$\left\langle \varphi_{\text{opt}}(x;\theta) \right\rangle_\theta = 0 \tag{11}$$

As a side remark we note that this property can be taken as a starting point for construction of the parameter estimator, and then one actually returns to the maximal likelihood method (cf. §16 in [6]). However, one then relinquishes the simple way to estimate errors as described above.

The key observation of the QOW method is that the deviation of the estimate from the optimal one (i.e. the one that corresponds to the Frechet-Rao-Cramer boundary) depends only quadratically on the deviation of the weight from the optimal one (for deviations which are not too large). This means that even non-ideal approximations for the weight would yield good estimates that are close to the optimum. One should also keep in mind that the deviation is an integral characteristic (see eq. (13) below), which gives one more freedom for constructing a quasi-optimal weight than is the case for point-wise function approximations.

The importance of this observation is seen from the fact that $\theta^*$ is unknown prior to the measurement, and it is impossible to point out the corresponding exact optimal weight. However, the quadratic (moreover, integral) smallness of deviations from the optimum often allows one to choose good, almost optimal weight without exact knowledge of $\theta^*$. Therein lies the crucial advantage of the QOW method.

The quadratic smallness can be expressed as follows. Suppose the weight $f(x)$ used in the method of moments is close to $\varphi_{\mathrm{opt}}(x;\theta^*)$: $f(x) = \varphi_{\mathrm{opt}}(x;\theta^*) + \delta f(x)$.
Consider the theoretical expression for the variance of the estimate for large $N$, then it is sufficient to consider the linear approximation for $h$:

$$\mathrm{Var}_{\mathrm{th}} \theta = \frac{\mathrm{Var}_{\mathrm{th}} f}{H(\theta^*)^2} \tag{12}$$

where $H(\theta) = \partial h(\theta)/\partial \theta$.

Expand this expression in $\delta f$, retain only quadratic terms, and denote $\overline{\delta f} = \delta f - \langle \delta f \rangle_{\mathrm{th}}$, then one obtains [2]:

$$\mathrm{Var}_{\mathrm{th}} \theta = \frac{1}{\langle \varphi_{\mathrm{opt}}^2 \rangle_{\mathrm{th}}} + \frac{1}{\langle \varphi_{\mathrm{opt}}^2 \rangle_{\mathrm{th}}^3} \left( \left\langle \left(\overline{\delta f}\right)^2 \right\rangle_{\mathrm{th}} \left\langle \varphi_{\mathrm{opt}}^2 \right\rangle_{\mathrm{th}} - \left\langle \varphi_{\mathrm{opt}} \overline{\delta f} \right\rangle_{\mathrm{th}}^2 \right), \tag{13}$$

where the quadratic term is non-negative due to the Schwartz inequality.

A convenient way to choose quasi-optimal weights is to take the optimal weight $\varphi_{\mathrm{opt}}(x;\theta_0)$ which corresponds to some a priori (e.g. known from previous experiments) value $\theta_0$ (thus assuming that $\theta_0 \approx \theta^*$). Then the deviation of the error from the theoretical lower bound would vanish quadratically as $(\theta_0 - \theta^*)^2$.

One should note the following, bearing in mind applications to problems which occur in accelerator measurements with a complicated event structure and computationally expensive theoretical Monte Carlo event generators: in view of the integral character of the above formulae, quasi-optimal weights could be chosen in the form of a rather crude, even piecewise-constant approximations.

Note that if the quasi-optimal weight is known as a function of $\theta$, then it becomes possible to improve the estimate iteratively. Indeed, let $\theta_1$ be the solution of the main equation (4). Then one can repeat the entire procedure taking the new value $\theta_1$ in place of $\theta_0$. This would yield a new solution $\theta_2$, and so on. It should be remembered, however, that the meaning of such iterations is not so much in the convergence of the estimates $\theta_i$ as in the minimization of the corresponding formal errors. We have called the errors formal because they are obtained following the general prescription of the basic method of moments, where the weight is independent of the sample. But in the iterative version the weight acquires a dependence on the sample via the estimates for $\theta$ obtained at the previous iterations, and then the true experimental error would be underestimated similar to how it is underestimated in the elementary case of estimation of the normal variance relative to the experimental mean, where in the simplest case of the normal distribution the corresponding correction is achieved via the Student distribution. There is no known analytical way for implementing such corrections in a general case. Therefore, we only note here that an iterative variant of the method must observe a balance between the improvement of the estimate due to iterations and the above effect of error underestimation. The eventual balance depends on a concrete experimental situation and must be studied concretely. This work only aims to understand the error estimation in the basic, non-iterative variant of the QOW method to provide a starting point for further improvements.

## 3. The case of insufficiently large $N$

In the simplest case the confidence interval is constructed via the standard deviation. This method is limited by the following conditions of applicability [1]:
1. The sampling distribution of the estimator becomes asymptotically normal as $N \to \infty$.
2. $N$ is sufficiently large for neglecting the deviations of the distribution from the normal one.

In order to study the case of insufficiently large $N$, rewrite the main equation (4) as follows:

$$\frac{1}{N}\sum_i \varphi_{\text{opt},i}(x_i;\theta_0) = \frac{1}{N}\sum_i \left\langle \varphi_{\text{opt},i}(x;\theta_0) \right\rangle_\theta \tag{14}$$

As was already mentioned, the standard deviation can be used as an error estimate only in the case of the asymptotically normal distribution of the estimator. On the left hand side, this is guaranteed by the central limit theorem [7], [8]. For small $N$, however, this approximation may not be valid. To clarify this point, recall the fundamental Berry-Esseen theorem [7]. Let $Y_i$ be independent random variables such that

$$\langle Y_i \rangle_{\text{th}} = \mu_i,$$
$$\text{Var}_{\text{th}} Y_i = \langle Y_i^2 \rangle_{\text{th}} - (\langle Y_i \rangle_{\text{th}})^2 = \sigma_i^2 < \infty, \tag{15}$$
$$\left\langle |Y_i - \mu_i|^3 \right\rangle_{\text{th}} = \rho_i < \infty$$

Let $F_N$ be the distribution function of the quantity $S_N = \dfrac{\sum_i (Y_i - \mu_i)}{\sqrt{\sum_i \sigma_i^2}}$. Then for any $y$, $N$ the following inequality is true:

$$|F_N(y) - \Phi(y)| \leq C\psi_N, \tag{16}$$

where $0.4097 \leq C < 0.5600$ (some concrete values are given in Table 1 [12]),

$$\psi_N = \left(\sum_i \sigma_i^2\right)^{-3/2} \sum_i \rho_i$$

and $\Phi(y)$ is the standard normal distribution.

| N | 1 | 2 | 3 | 4 | 5 | 6 | 7 | 8 | 9 | 10 |
|---|---|---|---|---|---|---|---|---|---|---|
| C(N) | 0.3704 | 0.4857 | 0.5111 | 0.5259 | 0.5356 | 0.5425 | 0.5476 | 0.5516 | 0.5547 | 0.5573 |

Table 1. Values for $C$ in Berry-Esseen inequality.

Denote $\Delta_{\text{BE},N} \equiv C\psi_N > 0$ and rewrite the inequality as follows:

$$\Phi(y) - \Delta_{\text{BE},N} \leq F_N(y) \leq \Phi(y) + \Delta_{\text{BE},N}, \qquad (17)$$

This yields pointwise estimates for the upper and lower bounds of the distribution.

When constructing the confidence interval, it is highly undesirable to overestimate its probability content, therefore consider the lower point:

$$F_N(y) \geq \Phi(y) - \Delta_{\text{BE},N} \qquad (18)$$

The distribution function is defined by

$$F_N(y) = \mathbb{P}[S_N \leq y] \qquad (19)$$

Therefore, with a given confidence level $\alpha$ for the theoretical mean $S^*$ of $S_N$ one obtains (the factor 2 comes from the fact that the bound on the absolute value is actually two-sided):

$$\mathbb{P}\Big[\big|S_N - S^*\big| \leq y_\alpha\Big] \geq \alpha - 2\Delta_{\text{BE},N} \qquad (20)$$

To err on the conservative side, consider the worst case, which corresponds to replacing the inequality with equality, and obtain:

$$\mathbb{P}\Big[\big|S_N - S^*\big| \leq y_\alpha\Big] = \alpha - 2\Delta_{\text{BE},N} \qquad (21)$$

In other words, the formally constructed confidence interval corresponds to a confidence level that is less than $\alpha$ by the amount $2\Delta_{\text{BE},N}$.

Since the quantity $\Delta_{\text{BE},N}$ which emerged from the Berry-Esseen theorem is independent of $\alpha$, one can implement the corresponding correction by simply rebuilding the confidence intervals with the confidence level $\alpha + 2\Delta_{\text{BE},N}$ instead of $\alpha$. Then

$$\mathbb{P}\Big[\big|S_N - S^*\big| \leq y_{\alpha+\Delta_{\text{BE},N}}\Big] = (\alpha + 2\Delta_{\text{BE},N}) - 2\Delta_{\text{BE},N} = \alpha \qquad (22)$$

To summarize, the confidence interval for the confidence level $\alpha$ for the theoretical mean $S^*$ in the case of small $N$ has the form:

$$S_N - y_{\alpha+2\Delta_{\text{BE},N}} \leq S^* \leq S_N + y_{\alpha+2\Delta_{\text{BE},N}} \qquad (23)$$

This confidence interval guarantees that it will cover the measured value $S^*$ with the probability *no less* than $\alpha$. Also note that the conservative choice made in eq. (21) may lead to $\alpha + 2\Delta_{BE,N} > 1$ in the case of large deviations from the asymptotic normality that cannot be treated as corrections anyway.

**Application to the QOW method.** Let us specify the quantities that participate in the Berry-Esseen construction, eq. (15), to the case of the QOW method:

$$Y_i = \varphi_{\text{opt},i}(x_i; \theta_0) \tag{24}$$

$$\mu(\theta) = \sum_i \left\langle \varphi_{\text{opt},i}(x; \theta_0) \right\rangle_\theta \tag{25}$$

$$\sigma^2(\theta) = \sum_i \left[ \left\langle \varphi^2_{\text{opt},i}(x; \theta_0) \right\rangle_\theta - \left( \left\langle \varphi_{\text{opt},i}(x; \theta_0) \right\rangle_\theta \right)^2 \right] \tag{26}$$

$$\rho(\theta) = \sum_i \left\langle \left| \varphi_{\text{opt},i}(x; \theta_0) - \left\langle \varphi_{\text{opt},i}(x; \theta_0) \right\rangle_\theta \right|^3 \right\rangle_\theta \tag{27}$$

$$\psi_N(\theta) = \left( \sum_i \sigma_i^2(\theta) \right)^{-3/2} \sum_i \rho_i(\theta) = \sigma^{-3}(\theta) \rho(\theta) \tag{28}$$

$$\Delta_{\text{BE},N}(\theta) = C \psi_N(\theta) > 0 \tag{29}$$

$$S_N(\theta) = \frac{\varphi_{\text{opt}}(x; \theta) - \mu(\theta)}{\sigma(\theta)} \tag{30}$$

Then the confidence interval for the weight is represented as follows:

$$\left| S_N(\theta_{\text{exp}}) - S_N(\theta^*) \right| \leq y_{\alpha + 2\Delta_{\text{BE},N}(\theta)} \tag{31}$$

There is still an ambiguity in the choice of the argument for $\Delta_{\text{BE},N}(\theta)$. It would be mathematically correct to take the value $\theta^*$ which, however, is unknown.
It is customary in such cases to choose $\theta_{\text{exp}}$ instead, especially that there is a safety margin built into the above construction due to the fact that the worst case bound was chosen from the Berry-Esseen inequality (see the reasoning for eq. (21)).

The aim here is to construct confidence intervals not for $S_N(\theta^*)$, but for the quantity $\varphi_{\text{opt}}(x; \theta^*)$ that occurs in the main equation of the method (4).
Therefore it is necessary to transform the interval (31) as follows:

$$\delta_-(x; \theta_{\text{exp}}, \theta^*) \leq \varphi_{\text{opt}}(x; \theta^*) \leq \delta_+(x; \theta_{\text{exp}}, \theta^*), \tag{32}$$

where

$$\delta_\pm(x; \theta_{\text{exp}}, \theta^*) = \varphi_{\text{opt}}(x; \theta_{\text{exp}}) + \left[ \mu(\theta^*) - \mu(\theta_{\text{exp}}) \frac{\sigma(\theta^*)}{\sigma(\theta_{\text{exp}})} \right] \pm \sigma(\theta^*) y_{\alpha + 2\Delta_{\text{BE},N}(\theta_{\text{exp}})} \tag{33}$$

As argued above, we substitute $\theta_{\text{exp}}$ here in place of the unknown $\theta^*$ and obtain the final interval for the weight:

$$\left| \varphi_{\text{opt}}(x; \theta_{\text{exp}}) - \varphi_{\text{opt}}(x; \theta^*) \right| \leq \sigma(\theta_{\text{exp}}) \times y_{\alpha + 2\Delta_{\text{BE},N}(\theta_{\text{exp}})} \tag{34}$$

From this, one obtains the following general interval for $\theta$ without invoking approximations for the function $h(\theta)$:

$$\theta_- \leq \theta^* \leq \theta_+ \tag{35}$$

$$\theta_{\pm} \triangleq h^{-1}\left(\varphi_{\text{opt}}\left(x;\theta_{\exp}\right) \pm \sigma\left(\theta_{\exp}\right) y_{\alpha+2\Delta_{\text{BE},N}\left(\theta_{\exp}\right)}\right) \qquad (36)$$

This formula assumes that $h(\theta)$ increases monotonically; in the case of decrease, the upper and lower bound should be swapped.

To summarize, the quantity $2\Delta_{\text{BE},N}\left(\theta_{\exp}\right)$ has a simple statistical meaning («the leakage of probability from the confidence interval») and can, therefore, be used as a convenient quantitative indicator of whether or not the normality assumption is valid for the distribution of the weight's mean.

## 4. The linear and quadratic approximations for $h(\theta)$

The geometric interpretation of the QOW method (see Fig. 1) allows one to immediately write down the confidence interval for $\theta^*$ with the confidence level $\alpha$, as follows:

$$\theta_{-} \leq \theta^* \leq \theta_{+}, \qquad (37)$$

where

$$\theta_{\pm} = h^{-1}\left(h\left(\theta_{\exp}\right) \pm \Delta f_{\alpha}\right), \qquad (38)$$

$h^{-1}(h)$ is the inverse function for $h(\theta)$, and the quantity $\Delta f_{\alpha} = y_{\alpha}\sigma_{\exp}$ was defined in eq. (7). For practical purposes (having in view the cases when the modelling and construction of $h(\theta)$ is expensive) it is convenient to consider the inversion of $h$ in eq. (32) in the linear and quadratic approximations, which is justified for large $N$ when the scatter of values of $h\left(\theta_{\exp}\right) \equiv \langle\varphi\rangle_{\exp}$ decreases with the growing $N$.

Start with the following quadratic approximation for $h(\theta)$ near $\theta_{\exp}$:

$$h(\theta) \approx h\left(\theta_{\exp}\right) + h'\left(\theta_{\exp}\right)\left(\theta - \theta_{\exp}\right) + \frac{1}{2}h''\left(\theta_{\exp}\right)\left(\theta - \theta_{\exp}\right)^2 \qquad (39)$$

After simple transformations one obtains:

$$h^{-1}(h) = \theta_{\exp} + \delta - \frac{1}{2}\gamma\delta^2, \qquad (40)$$

where

$$\gamma = \frac{h''\left(\theta_{\exp}\right)}{h'\left(\theta_{\exp}\right)}, \qquad \delta = \frac{h - h\left(\theta_{\exp}\right)}{h'\left(\theta_{\exp}\right)} \qquad (41)$$

Note that $\gamma = 0$ corresponds to the linear case. Substituting eq. (40) into the main formula (37), one obtains:

$$\left|\theta_{\exp} - \theta^* - \sigma_{\text{nlin}}\right| \leq \sigma_{\text{lin}} \qquad (42)$$

where

$$\sigma_{\text{lin}} \equiv \frac{\Delta f_{\alpha}}{h'\left(\theta_{\exp}\right)} \qquad (43)$$

$$\sigma_{\text{nlin}} \equiv \tfrac{1}{2}\gamma\,\sigma_{\text{lin}}^2 \qquad (44)$$

It is seen that neglecting quadratic terms would result in a systematic measurement error in the form of an overall shift of the confidence interval by $\sigma_{\text{nlin}}$.

The second effect of non-linearity is a possible distortion of the probability content of the confidence interval in the linear approximation compared with the quadratic case. In order to obtain a quantitative estimate, we use expression (42) to form the difference of probability contents of the two confidence intervals:

$$\Delta \mathbb{P} = \mathbb{P}\Big(\big|\theta_{\exp} - \theta^* - \sigma_{\text{nlin}}\big| \leq \sigma_{\text{lin}}\Big) - \mathbb{P}\Big(\big|\theta_{\exp} - \theta^*\big| \leq \sigma_{\text{lin}}\Big) \tag{45}$$

Introduce an intermediate quantity, namely, the distribution function $\Phi(\theta)$ of the estimator (it will drop from the final answer). Then the expression (45) can be rewritten as follows:

$$\Delta \mathbb{P} = \Big[\Phi\big(\theta_{\text{adj}} + \sigma_{\text{lin}}\big) - \Phi\big(\theta_{\text{adj}} - \sigma_{\text{lin}}\big)\Big] - \Big[\Phi\big(\theta_{\exp} + \sigma_{\text{lin}}\big) - \Phi\big(\theta_{\exp} - \sigma_{\text{lin}}\big)\Big] \tag{46}$$

where $\theta_{\text{adj}} \equiv \theta_{\exp} - \sigma_{\text{nlin}}$ (an adjusted $\theta_{\exp}$). Regroup the terms and obtain:

$$\Delta \mathbb{P} = \Big[\Phi\big(\theta_{\text{adj}} + \sigma_{\text{lin}}\big) - \Phi\big(\theta_{\exp} + \sigma_{\text{lin}}\big)\Big] - \Big[\Phi\big(\theta_{\text{adj}} - \sigma_{\text{lin}}\big) - \Phi\big(\theta_{\exp} - \sigma_{\text{lin}}\big)\Big] \tag{47}$$

In practice one works with confidence intervals with confidence levels that are close to 1 (100%). Then the variation of $\Phi$ in the region of interest is small, see Fig. 2.

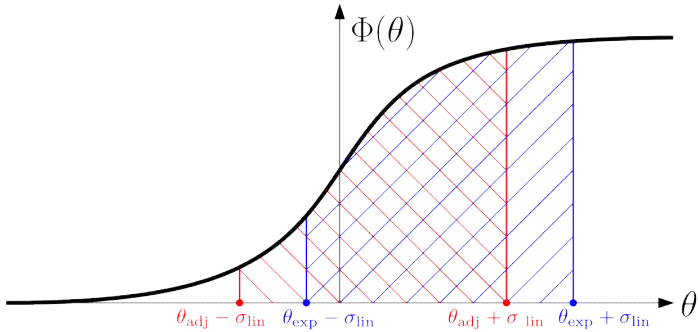

Fig. 2. Distribution function of $\theta_{\exp}$

Therefore, one can use approximations for each of the terms in the square brackets:

$$\Phi(\theta + \Delta) - \Phi(\theta) = \Phi'(\theta)\Delta = \pi_1(\theta)\Delta, \tag{48}$$

where $\pi_1(\theta)$ is the probability density for the estimator. Then eq. (45) becomes

$$\Delta \mathbb{P} = \sigma_{\text{nlin}} \times \Big[\pi_1\big(\theta_{\exp} + \sigma_{\text{lin}}\big) - \pi_1\big(\theta_{\exp} - \sigma_{\text{lin}}\big)\Big] \tag{49}$$

From the main equation of the QOW method, eq. (14), one can see that $\theta_{\exp}$ is connected to the empirical mean of the weight $\langle\varphi\rangle_{\exp}$. The distribution of $\langle\varphi\rangle_{\exp}$ is assumed to be normal (the corrections for non-normality in the sub-asymptotic case have been considered in sec. 2). Therefore, for a direct calculation it is more convenient to use the probability density of the empirical weight which we denote as $\pi_2(h)$ (the notation $h$ for the argument instead of the cumbersome $\langle\varphi\rangle_{\exp}$ is motivated by eq. (4)). Note that $\pi_2(h)$ is a normal distribution in accordance with the assumptions here; its parameters are discussed below. Using eq. (14), one can write (this is in fact a transformation of the probability measure under a mapping)

$$\pi_2(h)\,dh = \pi_1(\theta)\,d\theta \;\Rightarrow\; \pi_1(\theta) = \pi_2(h(\theta))\frac{dh}{d\theta} \qquad (50)$$

Finally, eq. (45) takes the form:

$$\Delta\mathbb{P} = \sigma_{\mathrm{nlin}} \times \left[\pi_2(h(\theta))\frac{dh}{d\theta}\right]_{\theta=\theta_{\mathrm{exp}}-\sigma_{\mathrm{lin}}}^{\theta=\theta_{\mathrm{exp}}+\sigma_{\mathrm{lin}}} \qquad (51)$$

This expression does not contain non-constructive quantities and can be implemented as a program, see sec. 5c.

$\Delta\mathbb{P}$ shows how the probability content differs between the linear approximations and the more accurate quadratic one. For instance, the value $\Delta\mathbb{P}=1\%$ would have to be viewed in the context of the target confidence level (e.g. 95%), and one would take further decisions based on that number depending on the concrete experimental context.

## 5. Examples

**5a. The Poisson case.** Let us work out eq. (51) for the case of Poisson distribution:

$$\pi(n;x,\theta) = \frac{\lambda(x,\theta)^n}{n!} e^{-\lambda(x,\theta)}, \qquad (52)$$

where $\lambda(x,\theta)$ is a known function that depends on the parameter $\theta$ to be estimated (the case of neutrino mass measurements in Tritium decays, cf. [5], [10]). Also known is the sample $\{(x_i, n_i)\}$ of size $N$ ($n_i$ is the count measured for the value $x_i$ of the control parameter $x$).

From the basic formula for the quasi-optimal weight, eq. (9) dropping the additive term (which is allowed by the method without loss of optimality, see sec. 2), one obtains:

$$\varphi_\theta(n;x,\theta) = n\frac{\partial \ln \lambda(x,\theta)}{\partial \theta} \equiv nA(x,\theta) \qquad (53)$$

Evaluate the main function of the QOW method $h(\theta)$. For a single control point $x_i$

$$h(x,\theta) = \sum_n nA(x,\theta_0)\frac{\lambda(x,\theta)^n}{n!}e^{-\lambda(x,\theta)} = \lambda(x,\theta)A(x,\theta_0) \qquad (54)$$

For the entire sample (cf. eq. (10)):

$$h(\theta) = \frac{1}{N}\sum_i h(x_i,\theta) = \frac{1}{N}\sum_i \lambda(x_i,\theta)A(x_i,\theta_0) \qquad (55)$$

Then the main equation of the QOW method becomes:

$$h(\theta) = \frac{1}{N}\sum_i n_i A(x_i,\theta_0) \qquad (56)$$

Numerically solving this for $\theta$ yields a point estimate $\theta_{\mathrm{exp}}$ for the unknown $\theta$.

One now has to evaluate eq. (51). There are three quantities whose values do not immediately follow from the basic QOW method: $\pi_2(h)$, $\sigma_{\mathrm{lin}}$, $\sigma_{\mathrm{nlin}}$.

$\pi_2(h)$ is the familiar normal distribution with the mean $\mu_2 = h(\theta_{\exp})$ and the variance $\sigma_2^2$ given by the following expression (see sec. 4 for explanations):

$$\sigma_2^2 = \frac{1}{N^2} \sum_i \lambda(x_i, \theta_{\exp}) [A(x_i, \theta_0)]^2 \tag{57}$$

To compute $\sigma_{\text{lin}}$, one first evaluates $\Delta f_\alpha$ according to eq. (7):

$$\Delta f_\alpha = y_\alpha \sigma_2. \tag{58}$$

Finally, in order to compute the loss of probability content due to the non-linearity, one substitutes eq. (58) into eqs. (43) and (44), and then use eq. (57) to construct $\pi_2(h)$.

Evaluation of the correction (29) that is due to the insufficiently large sample is essentially reduced to computation of $\psi_N(\theta)$ given by the following expression

$$\psi_N(\theta_{\exp}) = \sigma_2^{-3/2} \rho(\theta_{\exp}) \tag{59}$$

where

$$\rho(\theta_{\exp}) = \sum_i \rho(x_i, \theta_{\exp}) \tag{60}$$

$$\rho(x, \theta_{\exp}) = \sum_n \left| [n - \lambda(x, \theta_{\exp})] \frac{A(x, \theta_0)}{N} \right|^3 \frac{\lambda(x, \theta_{\exp})^n}{n!} e^{-\lambda(x, \theta_{\exp})} \tag{61}$$

The summation in (61) is to be done numerically.

$\sigma_{\text{nlin}}^*$ is computed from $\sigma_{\text{lin}}^*$ and the nonlinearity coefficient $\gamma$ (eq. (41)):

$$\gamma = \frac{h''(\theta_{\exp})}{h'(\theta_{\exp})} = \frac{\sum_i \lambda''(x_i, \theta_{\exp}) A(x_i, \theta_0)}{\sum_i \lambda'(x_i, \theta_{\exp}) A(x_i, \theta_0)} \tag{62}$$

Computer implementation of all the above is completely straightforward.

**5b. The case of the normal distribution.** The reasoning here follows the general scheme and is similar to the Poisson case with the integer count $n$ replaced by a continuous measurement $y$, and $\lambda(x, \theta)$ replaced by the pair $\mu(x, \theta)$, $\sigma(x)$:

$$\pi(y; x, \theta) = \frac{1}{\sqrt{2\pi} \sigma(x)} \exp\left(-\frac{(y - \mu(x, \theta))^2}{2\sigma^2(x)}\right), \tag{63}$$

where $\mu(x, \theta)$ is a known function that depends on the parameter $\theta$ to be estimated. The variance $\sigma^2(x)$ is assumed to be known for each value $x_i$ of the control parameter $x$ and to be independent of $\theta$. (If $\sigma^2(x)$ does depend on $\theta$, one has to take this into account when evaluating the derivatives, starting with the formula (64) that would become quadratic in $y$. It would simply make the formulae more cumbersome but is not a matter of principle.) Also known is the sample $\{(x_i, y_i)\}$ of size $N$.

The quasi-optimal weight is as follows:

$$\varphi_\theta(y; x, \theta) = y \frac{\partial \mu(x, \theta)}{\partial \theta} \frac{1}{\sigma^2(x)} \equiv y \times A(x, \theta) \tag{64}$$

The main function of the QOW method for a single control parameter $x_i$

$$h(x,\theta) = \mu(x,\theta) A(x,\theta_0), \qquad (65)$$

and for the entire sample:

$$h(\theta) = \frac{1}{N}\sum_i h(x_i,\theta) = \frac{1}{N}\sum_i \mu(x_i,\theta) A(x_i,\theta_0) \qquad (66)$$

The main equation of the QOW method is:

$$h(\theta) = \frac{1}{N}\sum_i y_i A(x_i,\theta_0) \qquad (67)$$

The auxiliary distribution $\pi_2(h)$ has the mean $\mu_2 = h(\theta_{\exp})$ and the variance $\sigma_2^2$ given by the following expression:

$$\sigma_2^2 = \frac{1}{N^2}\sum_i \sigma^2(x_i)\left[A(x_i,\theta_0)\right]^2 \qquad (68)$$

The Berry-Esseen correction is irrelevant here because the normal distribution is infinitely divisible so that the sum of normally distributed random variables is automatically normally distributed.

Finally, the formula for the nonlinearity coefficient $\gamma$ is the same as for the Poisson case (62) with $\lambda$ replaced by $\mu$.

**5c. Numerical example.** The numerical Monte Carlo test presented below illustrates how the above analytical formulae may work. The model is remotely reminiscent of, but much simpler than, what one deals with in the neutrino mass measurement experiments [4], [10]: the Poisson case (sec. 5a) with the following spectrum:

$$\lambda(x,\theta) = \begin{cases} 1 + (x-\theta)^2, & x > \theta \\ 1, & x \leq \theta \end{cases}$$

We used 10 uniformly distributed control points $x_i = 0, 2, ..18$, $i = 1,..10$. The "unknown" model value being measured is $\theta^* = 13$. We used 1 000 000 samples.

Straightforward calculations give the following results for the Berry-Esseen probability loss $\Delta_{BE,N}$ (see the analytical expression eq. (29)), the overall shift of the confidence interval $\sigma_{\text{nlin}}$ due to the non-linearity (see eq. (44)) together with a distortion of the probability content of the uncorrected confidence interval due to the nonlinearity $\Delta\mathbb{P}$ (see eq. (51)):

$$\Delta_{BE,N}(\theta_{\exp}) = 0.027; \quad \sigma_{\text{nlin}}^{\text{th}} = -0.06; \quad \Delta\mathbb{P} = -0.003 \qquad (69)$$

The Berry-Esseen correction was evaluated using the most conservative value of the constant $C = 0.56$, and the corresponding probability loss was then estimated as 5.4%. The probability distortion estimated using the quadratic approximation (see eq. (40)) turned out to be 0.3%. The resulting CL=90% interval for $\theta^*$ is:

$$\theta^* = 12.90 + 0.06 \pm 0.53 \qquad (70)$$

This includes the systematic overall nonlinear shift $0.06$, therefore the point estimate $\theta_{\exp} = 12.9$ is not in the center of the confidence interval, which may seem unusual. The effect of nonlinearity is clearly visible in Fig. 1.

In the Monte Carlo modelling, the confidence interval for the linear approximation (eq. (42) with $\sigma_{\text{nlin}} = 0$) covers the exact $\theta^*$ with probability 90.3%. In the quadratic approximation ($\sigma_{\text{nlin}} \neq 0$), the probability content becomes 90.2%, which is consistent with the negative $\Delta \mathbb{P}$ in eq. (69). For the exact confidence interval obtained using the inverse function $h(\theta)$, the probability content is 89.8%, which is less than the target 90% due to the Berry-Esseen effect. The Berry-Esseen correction to the confidence interval increases its probability content to 94.7%, which is conservatively larger than the target 90%, as expected. Note that the better results for the linear and quadratic approximations here is an example-specific fluctuation, the more significant number being the magnitude of the effect.

An interesting feature of this example is that the two corrections work in opposite directions, which is clearly seen in the two figures below: the Berry-Esseen correction reflects the fact that the histogram in the first Fig.3. leaks to the left from under the Gaussian shape, whereas the nonlinear correction, not knowing about the effect, attempts to preserve the probability content of the Gaussian confidence interval and map the result to the $\theta$ axis. The two figures below illustrate the two effects.

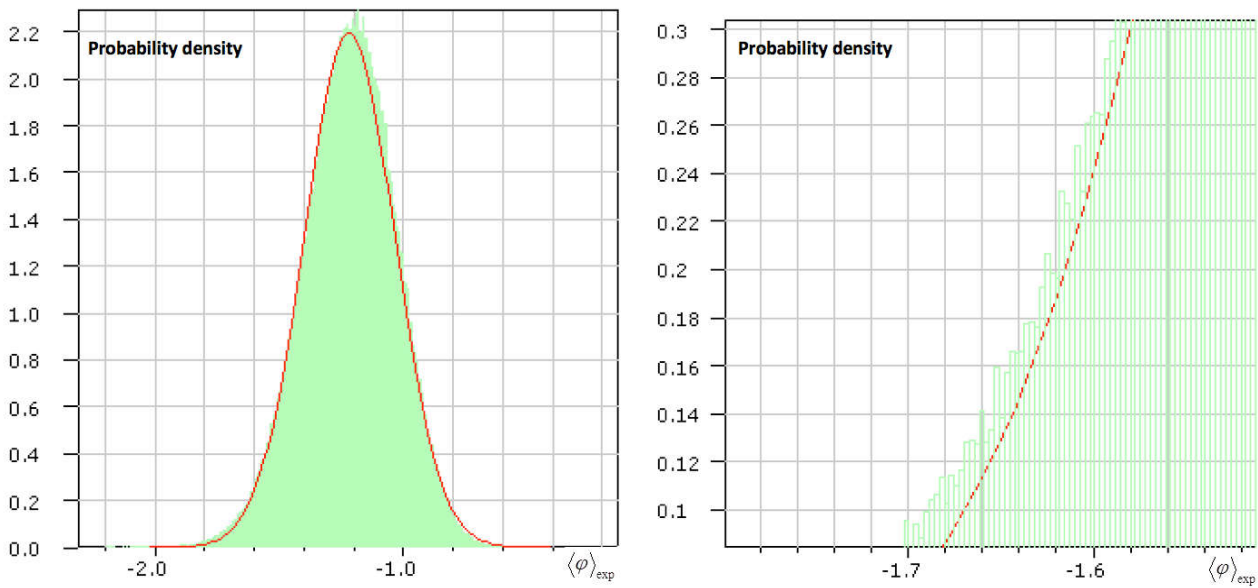

Fig. 3. The histogram of the average weight distribution, sec. 4, together with the corresponding asymptotic Gaussian. The Berry-Esseen correction to the confidence level attempts to (conservatively) estimate the probability leakage to the left of the Gaussian.

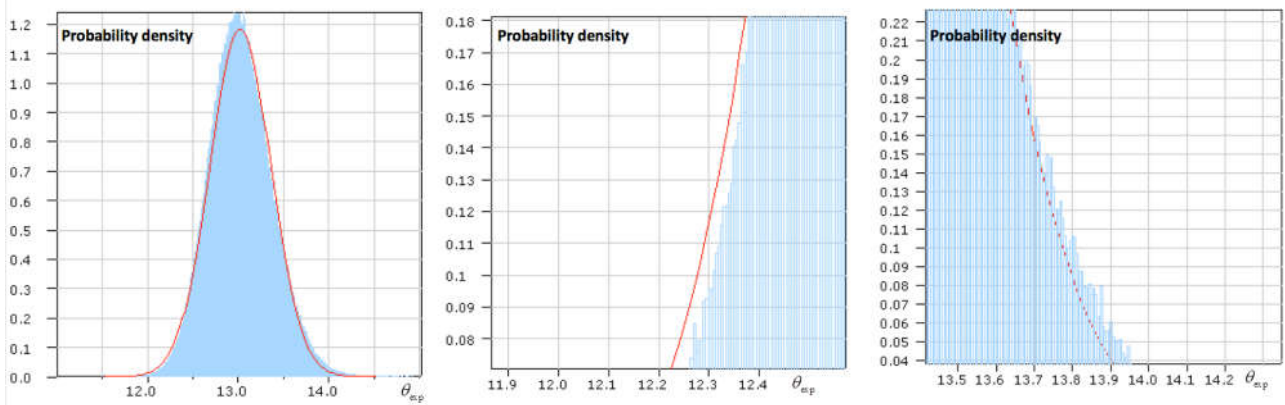

Fig. 4. The Monte Carlo distribution of the point estimate $\theta_{\rm exp}$. Also shown is the normal distribution that corresponds to the asymptotic large-$N$ assumption. A mismatch is seen, and it is this mismatch that the nonlinear correction $\sigma_{\rm nlin}$ attempts to correct for.

## 6. Conclusions

To summarize, the QOW method inherits the analytical transparency of the elementary method of moments and allows to explore, in much detail, the errors in the problem of parameter estimation. Constructive prescriptions emerge that allow a straightforward computer implementation. Eq. (29) is key for detecting a situation with insufficiently large $N$, eq. (51) takes into account the non-linearity. A concrete example of such errors evaluation is presented in sec. 5c. The obtained formulae are immediately applicable to situations with one parameter singled out on physical grounds (e.g. the neutrino mass [4], [10]), whose confidence interval is extracted from the marginal distribution. An important problem yet to be explored is the extension of the obtained results to the iterative variants of the QOW method (see the comments at the end of sec. 2).

**Acknowledgments**

We thank A.N. Titov (Institute for Nuclear Research RAS and KATRIN) for useful discussions and I.A. Denisov (Siberian Federal University and MOLPIT) for sharing a piece of software. This work was supported by R.Sh. Menyashev.